\documentclass[aps,pra,twocolumn,superscriptaddress]{revtex4-2}
\usepackage{graphicx}    
\usepackage{bbold}
\usepackage{dcolumn}    
\usepackage{amssymb}   
\usepackage{amsmath}    
\usepackage{amsthm}
\usepackage{commath}   
\usepackage{subfigure}    
\usepackage{braket}
\usepackage{natbib}
\usepackage{float}
\usepackage{color}
\usepackage{siunitx}
\usepackage[colorlinks=true, allcolors=blue]{hyperref}
\usepackage{hyphenat}

\usepackage[normalem]{ulem}

\usepackage{MnSymbol}
\usepackage{array}
\usepackage{dsfont}
\usepackage{physics}
\usepackage{empheq}
\usepackage{makecell}
\usepackage{mathtools}
\usepackage{multirow}
\usepackage{titletoc}

\newcommand{\up}{\Uparrow}
\newcommand{\down}{\Downarrow}
\newcommand{\Hamil}{\hat{H}}
\newcommand{\Hamilo}{\hat{H}_0}
\newcommand{\Hamilp}{\hat{H}'}
\newcommand{\Op}[1]{\hat{\sigma}_{#1}}

\newcommand{\thx}{\thanks{Equal contribution.}}

\newcommand{\pme}{\affiliation{Pritzker School of Molecular Engineering, University of Chicago, Chicago, Illinois 60637, USA}}
\newcommand{\physics}{\affiliation{Department of Physics, University of Chicago, Chicago, IL 60637, USA}}

\newcommand{\JFI}{\affiliation{James Franck Institute, University of Chicago, Chicago, IL 60637, USA}}

\graphicspath{{Figures/}}

\titlecontents{section}[0pt]
  {\small}
  {\contentslabel{2.2em}}
  {\hspace*{-2.2em}}     
  {\titlerule*[.5pc]{.}\contentspage}

\titlecontents{subsection}[2.2em]
  {\small}
  {\contentslabel{3em}}
  {}
  {\titlerule*[.5pc]{.}\contentspage}

\begin{document}

\title{Hybrid Device Design for a Pump-Free Microwave–Optical Bell-Pair Source}

\author{Jaesung Heo}\email{jheo@uchicago.edu}\thx\pme

\author{Fangxin Li}\email{fangxinli@uchicago.edu}\thx\physics

\author{Zhaoyou Wang}\pme

\author{Andrew P. Higginbotham}\physics\JFI

\author{Alexander A. High}\pme

\author{Liang Jiang}\email{liangjiang@uchicago.edu}\pme

\date{\today}

\begin{abstract}

Hybrid quantum devices offer a route to combine complementary functionalities of disparate physical systems within a single architecture. Extending this concept to a three-way interface among optical, microwave, and single-spin degrees of freedom is particularly challenging because of the weak magnetic interaction between an individual spin and a microwave photon. Here, we propose a hybrid device architecture that integrates a single diamond color center with a superconducting parallel-plate microwave resonator and a photonic-crystal cavity. The parallel-plate geometry drastically reduces the resonator impedance and enhances the single-spin microwave coupling rate by an order of magnitude. The enhanced interaction allows the optical cavity to be spatially separated from the superconducting structure while retaining appreciable microwave coupling and optical cooperativity above 1. For both NV$^-$ and ${}^{117}$SnV$^-$ centers, numerical simulations predict microwave coupling rates in the kilohertz range together with optical cooperativity exceeding 1. This architecture provides a promising route toward single-spin microwave-optical hybrid quantum devices enabling functionalities such as pump-free microwave-optical Bell-pair generation.
\end{abstract}

\maketitle

\section{Introduction}
Hybrid quantum architectures provide a route to combine the complementary advantages of different physical systems, enabling functionalities that are difficult to realize within individual platforms \cite{Xiang2013,Kurizki2015,Clerk2020}. For example, optical photons provide low-loss channels for communication \cite{Kimble2008,Takesue2015,Chen2021}, spins can serve as long-lived quantum memories \cite{Awschalom2018, Abobeih2018}, and superconducting circuits offer strong nonlinearities and rapid quantum processing \cite{Krantz2019}. Interfaces between these degrees of freedom have consequently enabled a broad range of functionalities, including spin-photon quantum-network nodes \cite{Nguyen2019PRL,Nguyen2019PRB,Stas2022,Knaut2024} and spin memories coupled to superconducting circuits \cite{Kubo2011,Grezes2014}.

Beyond pairwise optical-spin and microwave-spin interfaces, simultaneously integrating optical, spin, and microwave degrees of freedom within a single node can enable broader functionality, including microwave-optical (M-O) transduction \cite{Rochman2023,Xie2025} and entanglement generation \cite{Kurokawa2022, MO-NODE}. This possibility is particularly natural for spin systems that possess both optical and microwave transitions, such as diamond color centers. Such interfaces could enable distributed superconducting quantum computing and entanglement distribution through optical networks \cite{Kimble2008, Lauk2020, Han2021}.

\begin{figure}[t]
\centering
\includegraphics[width=0.95\linewidth]{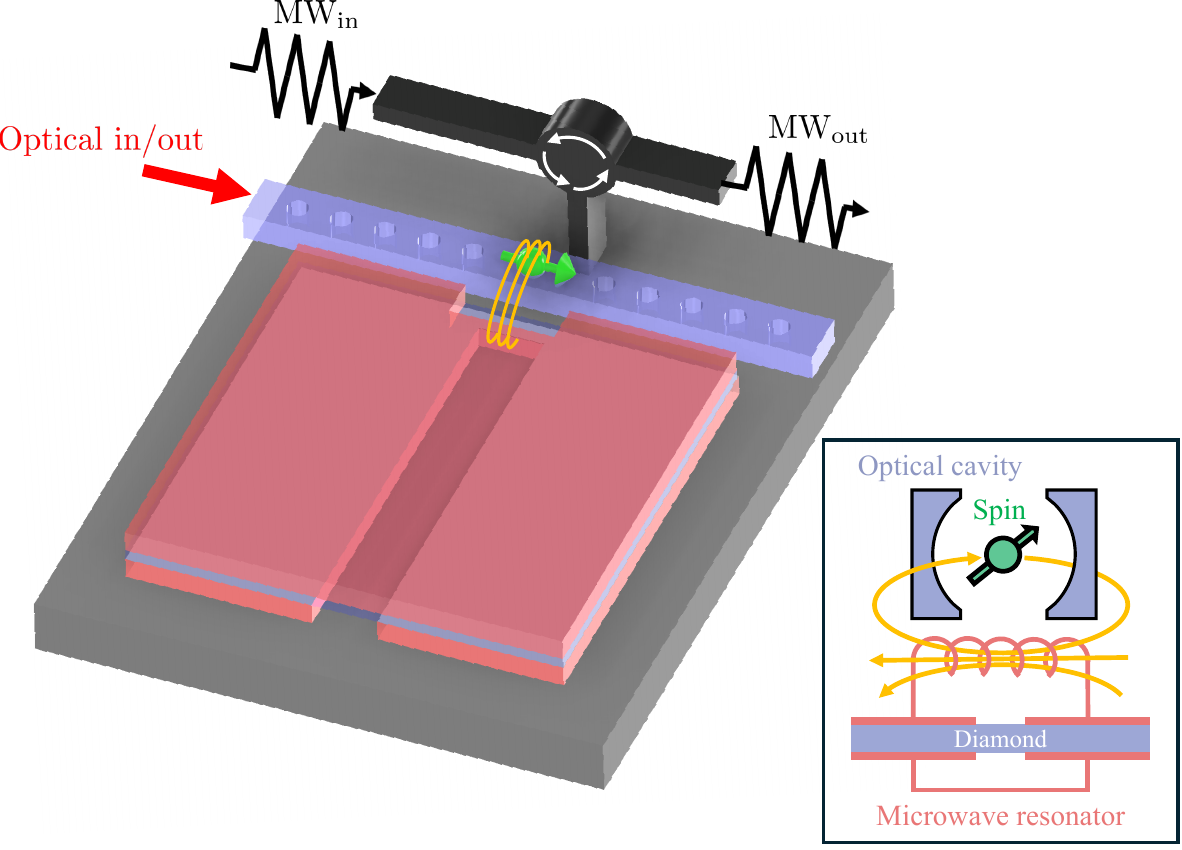}
\caption{Schematic of the hybrid microwave-resonator–optical-cavity device for pump-free microwave-optical Bell pair generation using a single diamond color center. The microwave resonator incorporates a diamond-sandwiched parallel-plate capacitor that enables strong coupling to a spin qubit in a one-dimensional photonic crystal cavity, thereby facilitating heralded generation of microwave–optical Bell pairs. \label{Fig1}}
\end{figure}

Implementing this three-way interface, however, introduces a nontrivial device co-design problem. To realize coherent functionality involving all three systems, such as pump-free generation of M-O entanglement, the spin must interact strongly with both microwave and optical modes at the single photon level \cite{MO-NODE}. Strong interaction with a microwave spin transition requires a large zero-point magnetic field at the spin position, favoring high zero-point currents and close proximity to the current-carrying structure. A high-quality optical interface, in contrast, requires the optical mode to remain sufficiently isolated from nearby conducting structures to avoid metal-induced absorption and cavity degradation \cite{Schietinger2009}. Collective enhancement can compensate for weak microscopic interactions, but it also introduces ensemble-specific complications, including inhomogeneous broadening and spatial mismatch between the microwave and optical modes \cite{Wesenberg2011,Kurucz2011,Julsgaard2013,Putz2014, Rochman2023, Xie2025}. A single-spin interface instead provides intrinsic microscopic nonlinearity and individual addressability while avoiding inhomogeneous broadening. These considerations motivate a hybrid architecture capable of achieving sufficiently strong interactions with a single spin.

Motivated by these considerations, we focus on a three-way hybrid device architecture based on a single diamond color center for a single-spin implementation of the pump-free M-O entanglement protocol MO-NODE (M-O No-Optical-Drive Entangler) \cite{MO-NODE}. Diamond color centers are promising spin emitters for such an architecture \cite{hanson2019,Rosenthal2023,Guo2023}. Recent experiments have demonstrated high-cooperativity coupling between single color centers and nanophotonic cavities, together with coherent microwave control of their spin states \cite{Sipahigil2016integrated,Nguyen2019PRB,Nguyen2019PRL}. These capabilities have enabled spin-photon entangling operations, long-lived quantum memories, multi-node quantum networking, and distributed blind quantum computation \cite{Stas2022,Knaut2024,Wei2025}.

By contrast, coupling diamond color center spins to quantized superconducting microwave modes has so far relied primarily on ensembles \cite{kubo2010strong,Amsuss2011}. At the single-spin level, the state-of-the-art direct magnetic coupling is in the few-kilohertz range, recently achieved using Er$^{3+}$ ions in the context of electron spin resonance spectroscopy \cite{wang2023single}. The challenge is even greater for the diamond color centers, whose effective gyromagnetic ratios are at least four times smaller than that of Er$^{3+}$. This makes enhancement of the microwave zero-point field particularly important for realizing a single diamond color center interface.

In this work, we address these requirements using a hybrid architecture based on a diamond membrane integrated with a superconducting parallel-plate resonator, as shown in Fig.~\ref{Fig1}. The parallel-plate geometry minimizes parasitic inductance in the capacitive section, thereby drastically reducing the characteristic impedance, with the resonator inductance predominantly provided by the inductive wire. The resulting enhancement of the zero-point magnetic field allows sufficient spatial separation between the superconducting structure and the optical mode while retaining appreciable coupling to a single spin. We consider the negatively charged nitrogen-vacancy center (NV$^-$) and the tin-vacancy center with strong nuclear hyperfine coupling (${}^{117}\mathrm{SnV}^-$) as representative diamond emitters. Numerical simulations predict single-photon microwave coupling rates $g_m/2\pi$ of a few kilohertz while maintaining optical cooperativity above 1. The resulting architecture provides a route toward resonant interaction between an individual diamond spin and microwave photons without sacrificing the optical interface, thereby providing the device functionality required for a single-spin realization of MO-NODE \cite{MO-NODE}.

\section{MO-NODE Application}

To demonstrate the functionality of the hybrid optical-spin-microwave design, we consider the following M-O entanglement generation protocol. Details of the protocol analysis can be found in Ref. \cite{MO-NODE}.

\begin{figure}[t!]
\centering
\includegraphics[width=\columnwidth]{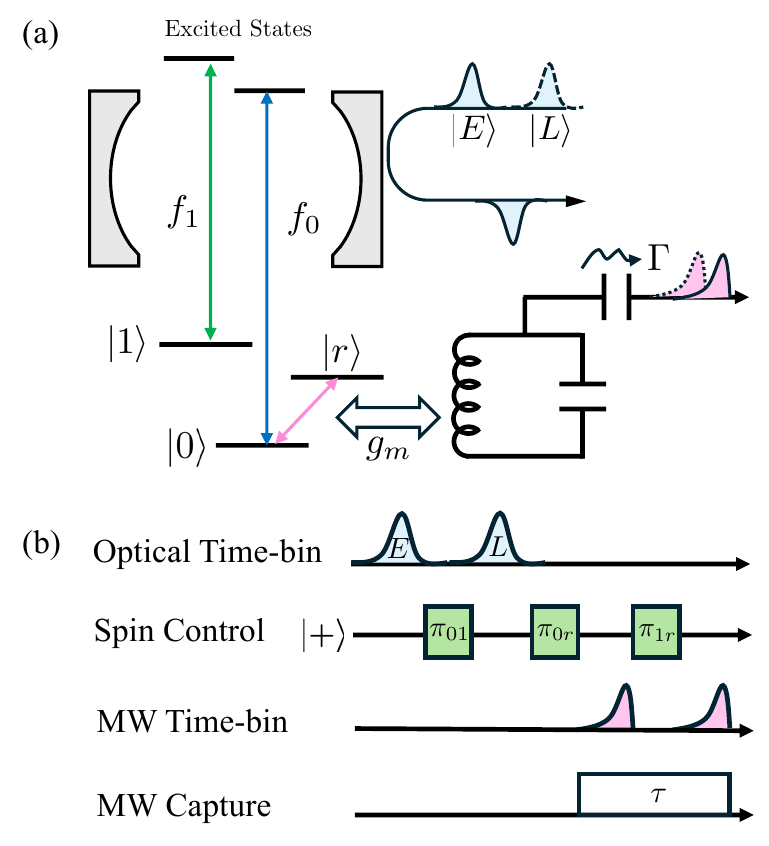}
\caption{(a) Illustration of the MO-NODE protocol with relevant states and parameters. The $\ket{0}$, $\ket{1}$ and $\ket{r}$ states are in the ground state manifold of a color center, which can be split by strain or magnetic field. These could be the spin-1 ground state of the NV$^-$, or the hyperfine ground states of ${}^{117}\text{SnV}^-$. The microwave readout transition $0-r$ is resonantly coupled to a LC microwave resonator. The optical transition to the excited state is coupled to an optical cavity. (b) Timed control sequences of the entanglement generation where $\pi_{0r}$ represents a $\pi$-rotation in the $\ket{0},\ket{r}$ subspace of the three-level system.
\label{Fig2}}
\end{figure}

Our entanglement generation protocol is realized in two steps. First, spin-photon entanglement is generated by spin-dependent optical reflectivity, a probabilistic high-fidelity protocol for diamond color centers in photonic crystal cavities \cite{PhysRevLett.92.127902, PhysRevX.4.031022,Nguyen2019PRB,PhysRevApplied.22.044013}. An optical photon in either time bin (early $\ket{E}$ or late $\ket{L}$) is reflected conditional on the spin state $\ket{0}$ or $\ket{1}$. The correlation between the time bins will be established by inserting a $\pi$-pulse on the spin, creating the final spin-photon Bell state $\ket{\Phi}_{\text{Photon,Spin}}=(\ket{E} \ket{1} + \ket{L}\ket{0})/\sqrt{2}$ with success probability $P_{\text{opt}}=0.5C^2/(C+1)^2$. Here, $C=\frac{g_o^2}{\kappa_o\gamma_o}$ is the spin-cavity cooperativity, where $g_o$ is the coupling between the relevant transition and the optical cavity; $\kappa_o$ is the loss rate of the optical cavity, and $\gamma_o$ is the optical decay rate. The prefactor $0.5$ is due to a non-reflective spin state, which discards photons half of the time. A series of experiments has demonstrated the protocol \cite{Nguyen2019PRB, Nguyen2019PRL, Stas2022, Knaut2024, Wei2025}, and the best-achievable Bell state fidelity has been shown to exceed $0.97$ \cite{Wei2025}.  

Next, we map the spin states to a microwave time-bin photon to create M-O Bell pairs. This can be done using an auxiliary readout level $\ket{r}$, whose decay to the ground state $\ket{0}$ is resonantly enhanced by the Purcell effect. The rate of the enhanced microwave emission is given by $\Gamma = 4g_m^2/\kappa_m$ for $g_m < \kappa_m$, where $g_m$ is the single-spin-microwave-photon coupling, and $\kappa_m$ is the linewidth of the microwave resonator. For an order-of-magnitude estimate, the highest achievable emission rate is limited by the spin-microwave coupling $\Gamma \sim g_m$. 

By sequentially bringing the spin state $\ket{0}$ and $\ket{1}$ to the readout level, we create the two time bins of the microwave photon. For a fixed duration $\tau$, the success probability of creating such a microwave state is $p_{\text{mw}} = 1-e^{-\Gamma\tau/2}$. 

The scheme exhibits a trade-off between the microwave success probability and state fidelity. As the time window $\tau$ increases, the microwave success probability $p_{\text{mw}}$ increases, but the output state fidelity decreases, as it is limited by the spin coherence time $T_2$: $\mathcal{F}_{\text{mw}} = \frac{1}{2}(1+e^{-\tau/T_2})$. Therefore, the performance of the scheme crucially relies on the spin-microwave coupling $g_m$, since a larger $g_m$ enables a fast microwave retrieval before the spin decoheres. 

In the full analysis of the in \cite{MO-NODE}, we also considered heralding the creation of M-O Bell pairs using an optical nondestructive setup and a three-level transmon performing a ``catching" protocol \cite{Niemietz2021, wenner_catching_2014, axline_-demand_2018, campagne-ibarcq_deterministic_2018, kurpiers_deterministic_2018,PhysRevApplied.12.044067}. Besides the measurement device and operation time limitations, the final heralding rate and the conditional fidelity of the full protocol are determined by the optical cooperativity $C$ and the spin-microwave coupling $g_m$, since higher $C$ and $g_m$ increase the success probability $P_{\text{opt}}$ and $p_{\text{mw}}$ of the protocol, and reduce the time during which the entangled state is exposed to dephasing in the spin environment. Therefore, to realize the functionality of the MO-NODE, simultaneously achieving high $C$ and $g_m$ forms the basis of the hardware design.

\begin{figure*}[t!]
\centering
\includegraphics[width=\linewidth]{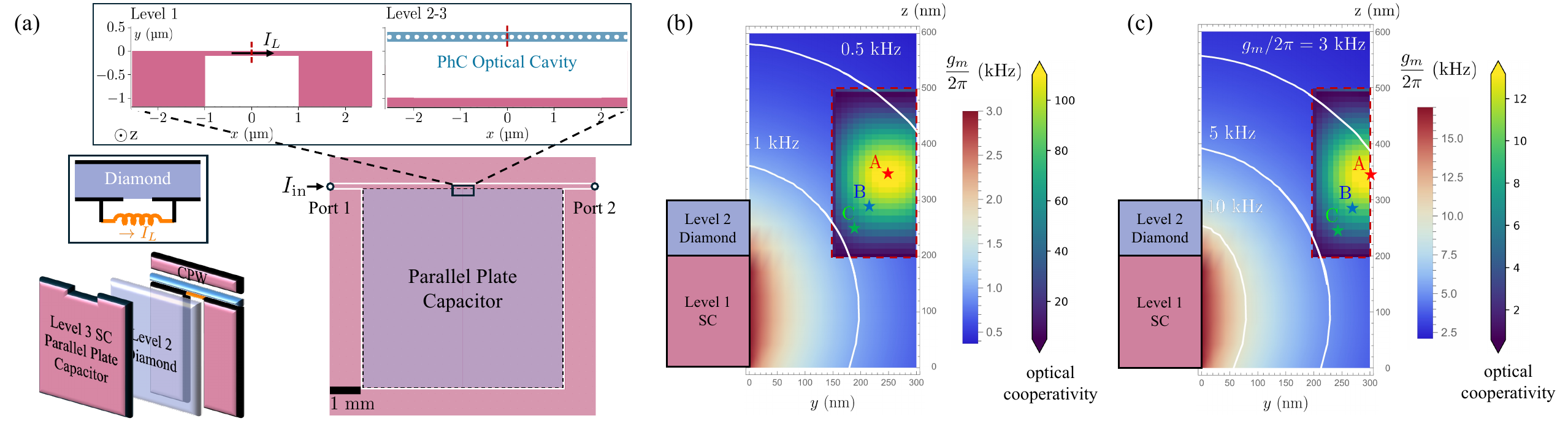}
\caption{(a) Hybrid optical-cavity-microwave-resonator design for strong coupling of a single diamond color center to both optical and microwave modes. The microwave resonator is modeled as a parallel LC circuit, with a coplanar waveguide (CPW) positioned above it for microwave signal input and readout. The resonator shown here is designed for the $^{117}$SnV$^-$ center. A one-dimensional photonic crystal cavity is positioned near the inductive-wire region of the resonator to achieve both strong microwave coupling and high optical cooperativity. (b,c) Simulation results evaluated at the center of the cavity (red dashed line in (a)) and the corresponding figures of merit for (b) the $^{117}$SnV$^-$ center and (c) the NV$^-$ center.}\label{Fig3}
\end{figure*}

\section{Microwave Strong Coupling}
We now turn our attention to a concrete implementation. We present a microwave resonator design that achieves a large coupling rate $g_m$ even at a distance where the optical cavity maintains high cooperativity $C$ despite metal-induced loss.

To maximize the coupling between the spin and the microwave resonator, a large magnetic zero-point fluctuation at the spin location is required. This field originates from the zero-point current fluctuations $\delta I$. For a lumped-element LC resonator with resonance frequency $\omega_R = 1/\sqrt{LC}$,
\begin{equation}\label{eq:di_z}
    \delta I = \omega_R \sqrt{\frac{\hbar}{2 Z}},
\end{equation}
where $Z = \sqrt{L/C}$ is the characteristic impedance of the resonator \cite{Haikka2017}. Because the resonance frequency is fixed by the spin transition, the design goal is to keep $\omega_R$ constant while minimizing $Z$.

The central element of our design is the use of a parallel-plate capacitor (Fig.~\ref{Fig1}). Previous spin–resonator coupling implementations have employed an interdigitated capacitor (IDC), where the capacitance originates from the electric field between interleaved fingers \cite{wang2023single, Eichler2017, Haikka2017,Ranjan2020}. Increasing the capacitance in this geometry requires many long fingers, which enlarge the device area and introduce significant stray inductance \cite{Eichler2017,wang2023single,Ranjan2020}, thus limiting impedance reduction. In contrast, the parallel-plate configuration provides orders of magnitude larger capacitance with minimal stray inductance, yielding a characteristic impedance up to three orders of magnitude lower than that of previously reported microwave resonators \cite{Ranjan2020, wang2023single, Eichler2017}.

The details of the resonator design are shown in Fig.~\ref{Fig3}(a). The microwave resonator consists of a three-level structure that effectively forms a parallel LC circuit. The bottom layer (level 1) comprises two large superconducting (SC) pads connected by a narrow inductive wire. The entire region is covered by a 100 nm thick single-crystal diamond layer (level 2) \cite{Guo2021,guo2024direct}. A large superconducting pad (level 3) covers the entire area except for the wire region, thereby suppressing image-current formation without significantly reducing the capacitance. Together, these layers form a parallel-plate capacitor that realizes the capacitive element of the LC circuit. The zero-point magnetic field fluctuation arises from the zero-point current fluctuations in the inductive wire (level 1). The target diamond defect center is embedded in the photonic crystal cavity positioned near the wire section of the resonator to ensure a large coupling rate $g_m$.

To send and read out microwave signals, a coplanar waveguide (CPW) is used with ports labeled 1 and 2. To evaluate the zero-point current fluctuations at the inductive wire $I_L$ in response to an input signal on the CPW, we analyzed the current-current susceptibility defined as $\chi_{II} = I_L/I_+$, where $I_+$ is the current incident at port 1. The zero-point fluctuations are then obtained as (see Appendix~\ref{spectral analysis} for details)
\begin{equation}
    {\delta I}^2 = \frac{\hbar \omega_R}{2 Z_0} \int_{-\infty}^\infty 2\ |\chi_{II}(\omega)|^2 \frac{d\omega}{2 \pi},
\end{equation}
where $Z_0=50~\Omega$ is the characteristic impedance of CPW.

\begin{table}
\begin{tabular}{|c|c|c|}
\hline  & NV$^-$\cite{doherty_nitrogen-vacancy_2013}  & ${}^{117}\text{SnV}^-$ \cite{harris2025high}\\
\hline Optical Transition Wavelength & $637$ nm  & $620$ nm \\
\hline Debye-Waller Factor ($\xi$) & 0.03 & 0.6  \\
\hline Optical Decay Rate $\gamma_o/2\pi$ & $6.1$ MHz  & $13.3$ MHz \\
\hline DC Magnetic Field $\vec{B}$ & 0.1-1 mT & 0 \\
\hline MW qubit frequency (GHz)  & 2.9 & 0.6 \\
\hline Magnetic Dipole & $\gamma_e/\sqrt{2}$  & $(0.4\gamma_e)/\sqrt{2} $\\
\hline Characteristic Impedance $Z$ & 0.12 $\Omega$ & 0.027 $\Omega$ \\
\hline Zero-Point Current Fluctuations $\delta I$ & 377 nA & 178 nA \\
\hline Capacitor Size (mm$^2$) & 0.95$\times$1.9 & 4.2$\times$8.5 \\
\hline 
\end{tabular}
\caption{Comparison of scheme parameters for the two representative color centers. The $\gamma_o/2\pi$ denotes the lifetime-limited HWHM. Optical linewidths approaching the lifetime-limited values have been demonstrated in diamond nanostructures \cite{orphal-kobin_optically_2023,trusheim_transform-limited_2020}. $\gamma_e$ is the gyromagnetic ratio of the electron. The magnetic dipole of the tin-vacancy center depends on the strain $\alpha$ in the system, and the spin-orbit coupling (see Appendix.~\ref{sec:color}). Parameters are used for calculating the optical cooperativity and spin microwave coupling strength.
}\label{defectproperty}
\end{table}

For concrete demonstration, we consider the NV$^-$ and the ${}^{117}\text{SnV}^-$ center as examples which show complementary characteristics. The properties of the two color centers are compared in Table~\ref{defectproperty} and elaborated in Appendix~\ref{sec:color}. The NV$^-$ requires at most a millitesla-scale DC magnetic field to lift the ground-state degeneracy. This is more than two orders of magnitude smaller than the DC magnetic fields under which Nb-based microwave resonators have already been demonstrated to operate while maintaining a high quality factor \cite{wang2023single, pscherer2026}. The ${}^{117}\text{SnV}^-$ allows for a zero-field protocol, as its ground state is already split by strain engineering and strong hyperfine interaction \cite{harris2025high}. Optically, the ${}^{117}\text{SnV}^-$ is much more efficient than the NV$^-$, which ensures a high spin-photon entanglement probability. This advantage in the ${}^{117}\text{SnV}^-$, however, is offset by a trade-off in the microwave domain. It has a reduced magnetic dipole moment compared to the NV$^-$, which results in lower microwave retrieval efficiency.

To understand the performance of our scheme, we simulate the device architecture designed for the color centers described in the previous paragraph. The inductive wire dominates the total inductance of the resonator, yielding $L=6.7$ pH, which is around two orders of magnitude smaller than the inductance of resonators employing an IDC design. The capacitor area is adjusted to match the microwave transition frequency of each color center, leading to different characteristic impedances and corresponding zero-point currents, as summarized in Table~\ref{defectproperty}.

\begin{table}
\begin{tabular}{|c|c|c|c|}
\hline
Color Center & Location & $C$ & $g_m/2\pi$ (kHz) \\
\hline
\multirow{3}{*}{${}^{117}\text{SnV}^-$} & A & 114 & 0.630 \\ \cline{2-4}
                        & B & 80.9 & 0.767 \\ \cline{2-4}
                        & C & 31.3 & 0.904 \\ \hline
\multirow{3}{*}{NV$^-$} & A & 13.6 & 3.18 \\ \cline{2-4}
                        & B & 9.65 & 3.78 \\ \cline{2-4}
                        & C & 3.73 & 4.35 \\ \hline
\end{tabular}
\caption{Simulation results of the optical cooperativity \(C\) and microwave coupling rate \(g_m\) evaluated at different spin positions in Fig.~\ref{Fig3}(b,c).}\label{C_g_comparison}
\end{table}

The resulting simulated coupling rate $g_m$ for the $^{117}$ SnV$^-$ and NV$^-$ centers are shown in Figs.~\ref{Fig3}(b) and \ref{Fig3}(c), respectively. The insets display the corresponding optical cooperativity where the cavity design follows the experimentally demonstrated thin-film diamond photonic-crystal platform of Ref.~\cite{ding2024high}. This platform has shown large optical quality factors, efficient photon-cavity coupling, spectrally stable single-emitter optical lines, and cavity-enhanced emission, making it a promising optical interface for our protocol. The optical cooperativity and microwave coupling rate for selected spin positions (marked A, B, and C) are shown in Table.~\ref{C_g_comparison}. These are chosen to be at least $30~\mathrm{nm}$ away from the nearest diamond surface, outside the near-surface regime where charge noise and spin decoherence are most severe~\cite{Myers2014}. Due to its larger magnetic dipole moment and higher qubit frequency, the NV$^-$ center achieves a microwave coupling rate $g_m$ that is 5.64 times larger than that of the $^{117}$ SnV$^-$ center. However, its weak zero-phonon line (ZPL) suppresses the optical cooperativity, which is 18 times lower than that of the $^{117}$ SnV$^-$ center, assuming the same optical-cavity-inductive-wire separation. Nevertheless, NV$^-$ can still achieve $C > 1$ and $g_m/2\pi > 1$ kHz simultaneously, provided that a small magnetic field is compatible with the experimental design. The ${}^{117}\text{SnV}^-$, despite having a smaller microwave coupling, remains a strong candidate for protocols requiring strictly zero-field operation. A comparison of different color centers, including the neutral silicon-vacancy center (SiV$^0$) as an
additional candidate, is discussed in Appendix.~\ref{sec:color}. 

\section{Discussion}
Diamond is a promising low-loss dielectric material for the experimental implementation of the parallel-plate capacitor design. Crystalline dielectrics are generally less susceptible to two-level-system loss than amorphous dielectrics \cite{OConnell2008}. For example, hBN has been shown to form high-quality interfaces with superconductors, supporting low loss tangents and high internal quality factors \cite{wang2022hexagonal}. Clean interfaces between diamond thin films with controlled dimensions and a wide range of materials containing oxide layers and complex structures have been realized by direct bonding, with covalent bonds formed across the interface \cite{Guo2021,guo2024direct,Ding2026}, suggesting a possible route toward low-loss diamond-superconductor interfaces. Niobium (Nb) is an attractive superconducting material due to its compatibility with oxide-mediated bonding approaches, its lattice constant being relatively close to that of diamond compared with previously demonstrated bonded materials \cite{guo2024direct}, and the high internal quality factors demonstrated in Nb-based microwave devices \cite{Verjauw2021,Kowsari2021}. To achieve $g\sim\kappa$, it is preferable for the total quality factor to exceed $10^5$, making a high internal quality factor a key experimental requirement.

Beyond our pump-free M-O Bell-pair generation protocol, this device platform may enable a broad range of applications, including compact superconducting circuits with long coherence and low crosstalk \cite{wang2022hexagonal}, strong coupling between spins and microwave photons \cite{wang2023single,pscherer2026}, and dispersive readout \cite{mamaev2026}. When combined with an optical cavity, microwave excitations mapped onto the spin state could be detected through cavity-enhanced optical readout, providing an optically assisted route to single-microwave-photon detection \cite{Anand2025}. Such a microwave-spin-optical interface could also be used for heralded remote entanglement between superconducting qubits through optical channels. Finally, by incorporating long-lived nuclear-spin levels as auxiliary storage states, the platform could be extended toward memory-integrated microwave-optical quantum transducers.

In summary, we propose a three-way hybrid device architecture that enables coherent interactions among an optical cavity, a single diamond color center, and a microwave resonator, providing the device functionality required for pump-free M-O entanglement generation. The architecture enables strong single-spin coupling to the microwave resonator while remaining compatible with a high-cooperativity optical cavity. Numerical simulations predict microwave coupling rates in the kilohertz range while maintaining optical cooperativity above unity, providing a pathway toward quantum information processing across disparate physical platforms.

\begin{acknowledgments}

We acknowledge helpful discussions with Benjamin Pingault, Xingyu Gao, Yang Shen, Judas Strayer, and Christopher Wang. We acknowledge support from the ARO(W911NF-23-1-0077), ARO MURI (W911NF-21-1-0325), AFOSR MURI (FA9550-21-1-0209, FA9550-23-1-0338), ONR MURI (N000142612102), DARPA (HR0011-24-9-0361), NSF (ERC-1941583, OMA-2137642, OSI-2326767, CCF-2312755, OSI-2426975). This material is based upon work supported by the U.S. Department of Energy, Office of Science, National Quantum Information Science Research Centers and Advanced Scientific Computing Research (ASCR) program under contract number DE-AC02-06CH11357 as part of the InterQnet quantum networking project.
 This work was completed with resources provided by the University of Chicago’s Research Computing Center. We acknowledge support from the Quantum Leap Challenge Institute for Hybrid Quantum Architectures and Networks (HQAN) (NSF OMA-2016136). We acknowledge help from the Sonnet Software Technical Support.  
\end{acknowledgments}


\begin{appendix}


\section{Methods}
\subsection{Microwave Resonator Modeling}\label{mc_simulation_imp}

We analyze our microwave resonator using an electromagnetic simulator, Sonnet v19.52 (Sonnet Software Inc.). For implementation, we considered a three-layer structure enclosed in a perfect conductor box. The bottom layer is a sapphire wafer with a thickness of $200~\mu\text{m}$. On top of it, a superconducting LC circuit along with a coplanar waveguide is patterned. This layer is set as a $100$ nm thick diamond. We used a conductor-backed coplanar waveguide structure to achieve $Z_0 = 50~\Omega$ \cite{CPW, Sahalos2019}. To realize the diamond-sandwiched parallel-plate capacitor structure, a superconducting metal layer covering the LC circuit area is drawn at the bottom of the third layer, which is air with a thickness of $10~\text{mm}$.

In designing the microwave resonator, it is essential to account for the kinetic inductance arising from the inertia of Cooper pairs in the superconducting film \cite{Annunziata2010, Haikka2017}. For the numerical modeling, we consider a 200 nm thick niobium film as a representative superconducting implementation, motivated by the high internal quality factors demonstrated in Nb-based superconducting circuits \cite{McRae2020, Kowsari2021}. Within BCS theory, the sheet kinetic inductance is given by \cite{Annunziata2010}
\begin{equation*}
    L_{K,\square} = \frac{R_{\square}\hbar}{\pi \Delta} \coth\left(\frac{\Delta}{2k_B T}\right),
\end{equation*}
where $R_{\square}$ is the normal-state sheet resistance, $\Delta$ is the superconducting energy gap, and $k_B$ is the Boltzmann constant. For 200 nm-thick niobium film, $R_{\square}=0.49\ \Omega$ \cite{Kowsari2021} and $\Delta=1.4$ meV \cite{Lap2021} yield $L_{K,\square}=0.073$ pH/sq.

\subsection{Analysis on Zero-Point Current Fluctuations and Finite-Element Simulation}\label{spectral analysis}

In this section, we describe our method to extract the zero-point current fluctuations directly from finite-element simulation results without relying on explicit mapping to a lumped-element circuit. The conventional approach for obtaining $\delta I$ is to infer the characteristic impedance $Z$ by fitting the simulated response to an equivalent lumped-element model. This is typically done by varying the participation of an inductive element or by comparing current ratios to deduce an equivalent lumped-element impedance $Z$ \cite{Eichler2017}. The problem is that microwave circuits do not operate in a strong lumped-element limit where components can be accurately approximated as ideal inductors or capacitors. For example, commonly used interdigitated capacitors introduce substantial stray inductance due to their extended finger geometry, altering the total circuit inductance and thereby leading to inaccurate estimates of the current fluctuations at the spin location. To avoid such issues, we introduce a method that computes the zero-point current fluctuations directly from the simulated current response and scattering parameters, eliminating the need for any lumped-element model extraction.

Let us denote the current at port 1 as $I_\text{in}$ and at the inductive wire as $I_L$ (Fig.~\ref{Fig3}(a)). From the Wiener–Khinchin theorem,
\begin{align}
    {\delta I}^2&=\left\langle I_L^2(0) \right\rangle = 2 \int_{-\infty}^\infty |\chi_{II}(\omega)|^2 S_{++}(\omega) \frac{d\omega}{2 \pi},\label{WK theorem}
\end{align}
where
\begin{equation}
    S_{++}(\omega) = \frac{\hbar \omega}{2 Z_0}\Theta(\omega)
\end{equation}
is the power spectral density of the incoming current fluctuations in port 1 \cite{Yurke1984}. The factor of two in Eq.~\ref{WK theorem} accounts for the zero-point fluctuations entering from port 2, which by symmetry are equal to those generated by those at port 1.

In the high quality factor limit, $\chi_{II}(\omega)$ is sharply peaked at the resonance frequency, enabling the frequency dependence of $S_{++}(\omega)$ to be neglected.
Eq.~\ref{WK theorem} can then be approximated as
\begin{equation}
    {\delta I}^2 = \frac{\hbar \omega_R}{2 Z_0} \int_{-\infty}^\infty 2\ |\chi_{II}(\omega)|^2 \Theta(\omega) \frac{d\omega}{2 \pi}.\label{I_L equation}
\end{equation}

To apply this analysis to the Sonnet simulation results, it is required to extract $I^+$ from $I_\text{in}$. The total current at port 1 is $I_\text{in} = I^+ - I^-$, where $I^-$ is the reflected current which can be obtained from the scattering matrix as $I^-=S_{11}I^+$ \cite{Pozar2012}. Therefore,
\begin{equation}
    I_\text{in} = (1-S_{11})I^+.
\end{equation}
Expressing the $\chi_{II}$ in terms of $I_\text{in}$, we obtain
\begin{equation}\label{current suscep}
    \chi_{II} = \frac{I_L}{I_\text{in}}(1-S_{11}).
\end{equation}
Substituting Eq.~\eqref{current suscep} into Eq.~\eqref{I_L equation} yields
\begin{equation}
    {\delta I}^2 = \frac{\hbar \omega_R}{2 Z_0} \int_{-\infty}^\infty 2\left|\frac{I_L(\omega)}{I_\text{in}(\omega)}\right|^2 \left|1-S_{11}\right|^2 \Theta(\omega) \frac{d\omega}{2 \pi}.
\end{equation}
This equation is useful because it gives the zero-point current fluctuations directly from simulation results.

\begin{figure}[t!]
\centering
\includegraphics[width=0.3\textwidth]{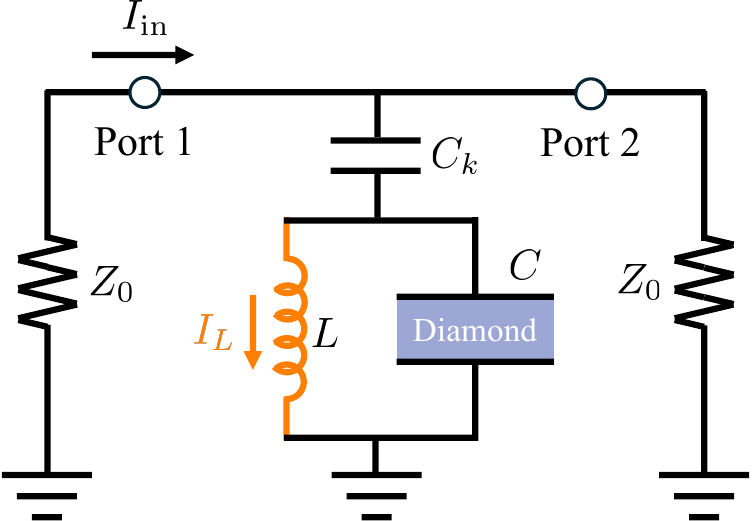}
\caption{Effective circuit diagram of the resonator coupled to a coplanar waveguide (CPW) transmission line. The circuit is conceptually equivalent to the model in Ref.~\cite{Eichler2017}.\label{effective circuit w transmission line}}
\end{figure}

As a check, we show that this expression leads to $\delta I=\omega_R\sqrt{\frac{\hbar}{2Z}}$ for an LC circuit with characteristic impedance $Z$. Fig.~\ref{effective circuit w transmission line} shows the effective circuit of the resonator coupled to the CPW, which is conceptually equivalent to the model in Ref.~\cite{Eichler2017}. Defining the impedance seen from port~1 as $Z_R$, we obtain
\begin{equation}
    \frac{1}{Z_R} = \frac{1}{Z_0} + \frac{1}{Z_k},
\end{equation}
where
\begin{equation*}
    Z_k=\frac{1}{i\omega C_k} + \frac{i\omega L}{1-\omega^2 LC}.
\end{equation*}
The reflection coefficient is $S_{11}$ \cite{Pozar2012},
\begin{equation}
    S_{11} = \frac{Z_{R}-Z_0}{Z_{R}+Z_0} \rightarrow 1-S_{11} = \frac{2 Z_0}{Z_{R}+Z_0}.
\end{equation}
The fraction of the input current $I_\text{in}$ that flows into the inductive wire ($I_L$) is determined by the impedance relation, given by
\begin{equation}
    \frac{I_L}{I_\text{in}} = \frac{Z_0}{Z_0+Z_k}\frac{1}{1-\omega^2 LC}.
\end{equation}
We now introduce the dimensionless variables
\[
Z' = \frac{Z_k}{Z_0},\quad 
\omega' = \frac{\omega}{\omega_R},\quad
C' = \omega_R Z_0 C_k,\quad
L' = \frac{\omega_R L}{Z_0},
\]
which yield
\begin{align*}
    &Z'=\frac{1}{i\omega' C'} + \frac{i\omega' L'}{1-\omega'^2}, \\
    &1-S_{11}= 1+\frac{1}{2Z'+1},\\
    &\frac{I_L}{I_\text{in}}= \frac{1}{1+Z'}\frac{1}{1-\omega'^2}.
\end{align*}
Thus, the current–current susceptibility becomes
\begin{equation*}
    \chi_{II}(\omega')= -\frac{2\omega' C'}{2i-\omega' C'-2i\omega'^2(1+L' C')+\omega'^3C'}.
\end{equation*}
Expanding $1/\chi_{II}$ for $\omega'$ about the point $1/\sqrt{1+L' C'}$ gives
\begin{align*}
    \frac{1}{\chi_{II}}=&\frac{L'C'}{2+2L'C'}\\
    &+\left(\frac{2i}{C'}+2iL'-\frac{1}{\sqrt{1+L'C'}}\right)\left(\omega'-\frac{1}{\sqrt{1+L'C'}} \right)\\ &+ \mathcal{O}\left[\left(\omega'-\frac{1}{\sqrt{1+L'C'}}\right)^2\right].
\end{align*}
In the limit $C'\rightarrow 0$,
\begin{align*}
    \chi_{II}(\omega')&\approx \frac{\frac{C'}{2+2L'C'}}{\frac{L'{C'}^2}{(2+2L'C')^2}+i\left(\omega'-\frac{1}{\sqrt{1+L'C'}}\right)}\\
    &\approx \frac{2}{L'C'}\frac{\frac{L'{C'}^2}{4}}{\frac{L'{C'}^2}{4}+i\left(\omega'-\frac{1}{\sqrt{1+L'C'}}\right)},
\end{align*}
where the last part is a Lorentzian form. The integral of the $|\chi_{II}(\omega')|^2$ is then,
\begin{equation*}
    \int_{-\infty}^\infty |\chi_{II}(\omega')|^2\frac{d\omega'}{2\pi}\approx\frac{1}{2L'}\rightarrow \omega_R\frac{Z_0}{2\omega_R L}=\frac{Z_0}{Z}\frac{\omega_R}{2},
\end{equation*}
where the final step restores the original frequency and inductance units.
Therefore, 
\begin{equation}
    \delta I^2 = \frac{\hbar{\omega_R}}{Z_0}\cdot\frac{Z_0}{Z}\frac{\omega_R}{2}={\omega_R}^2\frac{\hbar}{2Z},
\end{equation}
which gives the zero-point current fluctuations of the resonator:
\begin{equation}
    \delta I=\omega_R\sqrt{\frac{\hbar}{2Z}}.
\end{equation}

\subsection{Microwave Circuit Parameter Analysis}

From the simulation data, we obtain $\omega_R$ and $\delta I$, which are both expressed in terms of circuit inductance $L$ and capacitance $C$. Therefore, it is possible to extract circuit parameters from the simulation data,
\begin{align}
    C &= \frac{1}{\omega_R Z} = \frac{2 {\delta I}^2}{\hbar {\omega_R}^3}, \\
    L &= \frac{Z}{\omega_R} = \frac{\hbar}{2}\frac{\omega_R}{{\delta I}^2}.
\end{align}

For the test circuit with two capacitor pads (1 mm by 2 mm) on level 1 connected by an inductive wire of length 250 µm and 5 µm wide with level 3 metal covering the wire region as well, $C = 503$ pF and $L = 15.48$ pH, which matches very well with the capacitance of the parallel-plate capacitor with dielectric constant $\epsilon_r = 5.76$, which is expected to be 510 pF. The inductance of a wire covered by a superconducting pad can be simulated in Sonnet by setting two ports on each side of inductive wire and applying basic inductance measurement model. This simulation predicts an inductance of 13.2 pH, which agrees well with the $L$ obtained. Changing the inductive wire geometry to 50 µm long and 5 µm wide, this LC circuit gives $C = 504$ pF and $L = 4.92$ pH, where the wire inductance is expected to be $2.64$ pH. These results show that our analysis agrees with the expected results and the circuit inductance is dominated by the wire inductance.

\begin{table}
\begin{tabular}{|c|c|c|}
\hline $L$ (pH) & $C$ (pF) & $C_\text{cal}$ (pF) \\
\hline 7.03 & 1.16 & 1.27 \\
\hline 6.69 & 5.03 &  5.10 \\
\hline 6.70 & 11.49 &  11.47 \\
\hline 6.75 & 20.55 & 20.40 \\
\hline 
\end{tabular}
\caption{Simulated circuit parameters ($L$ and $C$) and calculated capacitance $C_\text{cal}$ for a microwave resonator with a fixed inductive wire ($2~\mu\mathrm{m} \times 0.1~\mu\mathrm{m}$) and varying capacitor area.}\label{wire scale comparison}
\end{table}

\begin{figure*}
    \centering
    \includegraphics[width=0.95\linewidth]{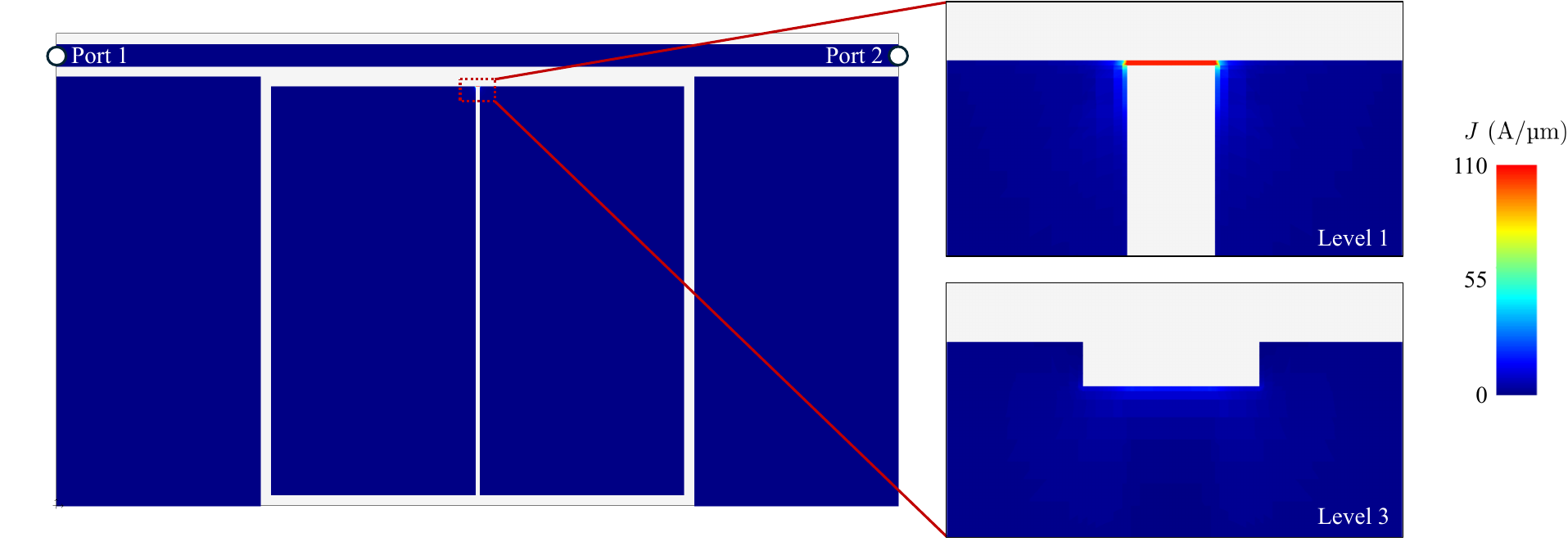}
    \caption{Current distribution plot of the circuit with 50 µm by 100 µm capacitor pad, where 1(0) V is given in port 1(2). Strong current confinement is observed only in the inductive wire region.}
    \label{fig:current_dist}
\end{figure*}

Next, we examine whether the stray inductance scales with the capacitor size. Using a fixed inductive wire geometry (length $2$ µm, width $0.1$ µm), Table~\ref{wire scale comparison} summarizes the simulated circuit parameters for increasing capacitor area. Here, $L$ and $C$ are obtained from Sonnet simulations, while $C_{\mathrm{cal}}$ is the capacitance estimated from the ideal parallel-plate capacitor expression. The results show that the inductance $L$ remains nearly constant even as the capacitance increases, demonstrating that the stray inductance does not scale with the capacitor size. Fig.~\ref{fig:current_dist} shows the simulated current distribution for the circuit with a 50 µm by 100 µm capacitor pad, where 1 V is applied to port 1 and 0 V to port 2. This clearly shows that the current is strongly concentrated in the inductive wire. These results confirm that the dominant contribution to the total inductance originates from the inductive wire rather than the capacitor geometry.

\subsection{Microwave Resonator–Spin Coupling Rate}
The microwave coupling rate is given by \cite{Haikka2017}
\begin{equation}
g_m = -\gamma_e \delta\mathbf{B}\cdot\langle 0|\mathbf{S}|r\rangle,
\end{equation}
where $\gamma_e/2\pi = 28~\mathrm{GHz/T}$ is the electron gyromagnetic ratio, $\delta\mathbf{B}$ denotes the zero-point magnetic field fluctuations, and $\langle 0|\mathbf{S}|1\rangle$ is the spin transition matrix element. The field $\delta\mathbf{B}$ is evaluated using the Biot–Savart law with the zero-point current fluctuations $\delta I$. The magnetic dipole moment $\gamma_e\langle 0|\mathbf{S}|1\rangle$ depends on the specific spin system, as summarized in Table~\ref{colortab}.

\subsection{Metal-Induced Optical Loss Simulation}

Our optical cavity design is based on the high quality factor photonic crystal (PhC) cavity demonstrated in Ref.~\cite{ding2024high}, where the PhC cavity is suspended in air to maximize vertical optical confinement and minimize substrate losses. In contrast, our device employs a substrate-supported geometry optimized for integration with the superconducting microwave circuit, as shown in Fig.~\ref{Fig3}. Specifically, the multilayer stack consists of a 200 µm thick sapphire substrate, a 200 nm thick niobium film, and a diamond membrane structured such that the niobium region retains 100 nm of diamond while the optical cavity region preserves 300 nm. A second 200 nm thick niobium layer is subsequently deposited atop the 100 nm diamond region. The overall PhC slab geometry is 300 nm thick and 200 nm wide.

To accommodate this geometry, we optimized the photonic crystal parameters for $\text{SnV}^-$ to $a = 0.1925$ µm, $r = 48~\mathrm{nm}$, and a minimum taper size of $a \times 0.8845$ in the cavity region, using quadratic tapering over six periods \cite{ding2024high}. The resulting cavity supports a TM-like mode, in contrast to the TE-like mode in Ref.~\cite{ding2024high}, thus minimizing evanescent field overlap with the nearby metal layer. Finite-difference time-domain simulations were performed using Tidy3D to evaluate the optical properties. Metal-induced losses were included by modeling the metal region as gold. Additional simulations using niobium yielded the same qualitative dependence on cavity-metal separation and did not alter the conclusions in the operating regime. When the cavity edge was positioned 500 nm from the metal edge, the simulated quality factor was $Q = 1.86 \times 10^{4}$, which corresponds to the intrinsic cavity intensity linewidth (FWHM) $\kappa_{o,i}/\pi=25.8$ GHz. For other cavity–metal separations $d$ (in nm), the corresponding values of $\kappa_{o,i}$ (in GHz) are $(d,\kappa_{o,i}/\pi)=(200, 36.5),\ (150, 77.7),\text{ and } (100, 230)$.

To evaluate the cooperativity of our optical cavity, we require three quantities: the cavity coupling rate $g_o$ to the selected ZPL transition, the cavity decay rate $\kappa_o$, and the emitter coherence decay rate $\gamma_o$. The cavity coupling rate is given by \cite{Reiserer2015}
\begin{align*}
    g_o(\vec{r})=\frac{\mu E}{\hbar} = \sqrt{\frac{\mu^2 \omega_o}{2\epsilon(\vec{r})\epsilon_0 \hbar V}}u(\vec{r}),
\end{align*}
where
\begin{align*}
    &u(\vec{r})=\sqrt{\frac{\epsilon(\vec{r})\left| E(\vec{r})\right|^2
    }
    {
    \text{max}\left[\epsilon(\vec{r})\left| E(\vec{r})\right|^2 \right]
    }},\\
    &V=\int u(\vec{r})^2 d^3 r.
\end{align*}
Here, $\mu$ is the electric dipole moment, $\omega_o$ is the optical cavity frequency, $\epsilon_0$ is the vacuum permittivity, $\epsilon(\vec{r})$ is cavity permittivity distribution, and $E(\vec{r})$ is the cavity electric-field profile. From the optical cavity simulations, we obtain $\epsilon(\vec{r})$, $E(\vec{r})$, and the cavity loss rate $\kappa_o$. We considered critical coupling, taking $\kappa_o=\kappa_{o,i} + \kappa_{o,e}\approx 2\kappa_{o,i}$, where $\kappa_{o,e}$ is the extrinsic decay rate. The spin-dependent quantities are the dipole moment $\mu$ and the optical coherence decay rate $\gamma_o$, given by 
\begin{align*}
    &\mu=\sqrt{\frac{3\pi\epsilon_0\hbar c^3\xi}{n\omega^3\tau_\text{rad}}}, \\
    &\gamma_o = \frac{1}{2\tau_\text{rad}}+\frac{\Gamma_{\text{nr}}}{2} + \gamma_\phi,
\end{align*}
where $c$ is the speed of light, $n$ is the refractive index (for diamond, $n=2.4$), $\omega$ is the zero phonon line (ZPL) transition frequency, $\tau_\text{rad}$ is the radiative lifetime, and $\xi$ is the Debye-Waller factor. The coherence decay rate $\gamma_o$ consists of radiative and non-radiative relaxation $\frac{1}{2\tau_\text{rad}}+\frac{\Gamma_{\text{nr}}}{2}$ and dephasing rate $\gamma_\phi$. Under the radiative, lifetime-limited assumptions $\gamma_o = 1/2\tau_\text{rad}$. Some experiments have shown to approach this life-time limited value~\cite{orphal-kobin_optically_2023, trusheim_transform-limited_2020}. The relevant definition of $\gamma_o$ for optical cooperativity is discussed in Sec.~\ref{app:optical-cooperativity}. The parameters used in our calculation are listed in Table~\ref{colortab}. The resulting optical cooperativity $C=g_o^2/\kappa_o\gamma_o$ is plotted in Fig.~\ref{Fig3}. 

\section{Scheme Compatible Color Centers}\label{sec:color}

In the main text, we demonstrate our scheme using the nitrogen-vacancy center and the tin-vacancy center. Our scheme can be generalized to other single defect centers provided that they satisfy the following requirements. For generating spin-photon entanglement, the system must demonstrate a robust spin initialization, well-resolved spin-dependent optical transitions, and microwave control over the spin levels. To realize our microwave readout scheme, the spin states must contain a qubit subspace and a gigahertz-range readout level that couples strongly to microwave. Moreover, the qubit subspace and the readout level must be connected with allowed transitions to enable microwave control. 

\begin{figure}[t!]
\centering
\includegraphics[width=0.9\columnwidth]{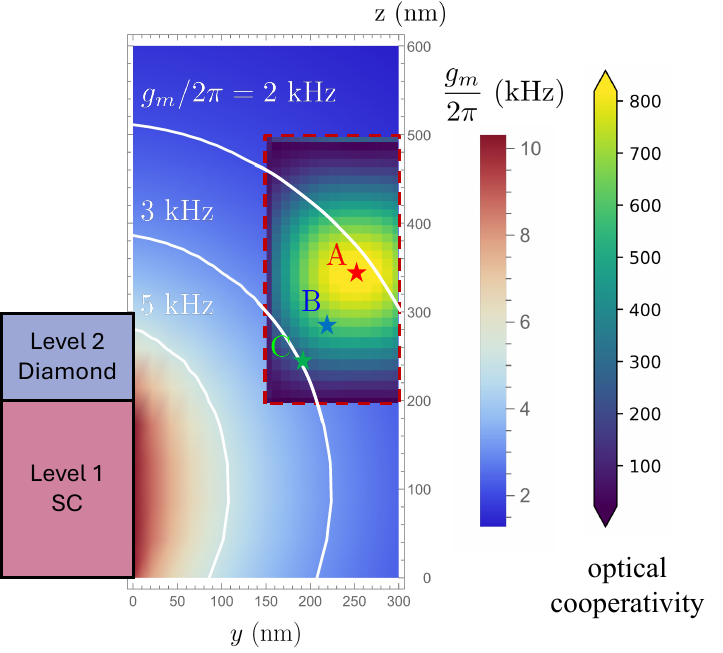}
\caption{Simulated optical cooperativity and microwave coupling rate of an SiV$^0$ center.
\label{neutral_SiV}}
\end{figure}

\begin{table}
\begin{tabular}{|c|c|c|c|}
\hline
Color Center & Location & $C$ & $g_m/2\pi$ (kHz) \\
\hline
\multirow{3}{*}{SiV$^0$} & A & 845 & 2.10 \\ \cline{2-4}
                         & B & 601 & 2.55 \\ \cline{2-4}
                         & C & 232 & 3.01 \\ \hline
\end{tabular}
\caption{Optical cooperativity $C$ and microwave coupling rate $g_m$ for SiV$^0$ at different locations shown in Fig.~\ref{neutral_SiV}.}
\label{fom SiV0}
\end{table}

Besides the demonstrated color defects in the main text, the neutral silicon-vacancy ($\text{SiV}^0$) center could also be a potential candidate \cite{Rose2018, Zhang2020}, although only ensemble spin resonance of $\text{SiV}^0$ has been demonstrated \cite{bhaskar_diamond_2021}. The simulated optical cooperativity and microwave coupling rate of an $\text{SiV}^0$ are shown in Fig.~\ref{neutral_SiV} and Table.~\ref{fom SiV0}. The properties of the three color centers are compared in Table~\ref{colortab}. The NV$^-$ and SiV$^0$ have a spin-1 ground state $m=\pm1,0$. Two of the levels ($\pm1$) are degenerate in the zero-field limit. To establish a long-lived qubit subspace, and strongly coupled readout level, the three levels need to be sufficiently split by a DC magnetic field. To create a mega-hertz level splitting between the $\pm1$ states, which makes the state decay negligible ($<10^{-3}$) during the readout, a small magnetic field ($0.1-1$mT) is needed. The advantage of the spin-1 ground state is that it enables strong spin-microwave coupling, with magnetic dipole moment given by $\gamma_e\bra{0}\vec{S}\ket{1}=\gamma_e/\sqrt{2}$ \cite{Haikka2017}. The ${}^{117}\text{SnV}^-$ ground state can be split by strain engineering and strong hyperfine interaction, as demonstrated in \cite{harris2025high}. Therefore, it allows the operation of the scheme in zero-field. However, due to the hybridization of electron-nuclear spin in the ground state, the ${}^{117}\text{SnV}^-$'s dipole moment is less than half of that of the NV$^-$ and SiV$^0$. This is shown by a perturbation calculation in the Appendix.~\ref{Perturbation}. The small magnetic dipole moment results in a low microwave retrieval efficiency. Moreover, due to the low transition frequency $\sim0.6$ GHz, the ${}^{117}\text{SnV}^-$ faces higher thermal noise and potential frequency mismatch when coupled to a transmon qubit, which may be resolved by operating the scheme at a lower cryogenic temperature and employing a flux tunable superconducting qubit. 

In terms of optical properties, the NV$^-$ has a very small ($\sim0.03$) Debye-Waller factor (D-W factor), compared with a D-W factor of 0.9 for SiV$^0$ and 0.6 for ${}^{117}\text{SnV}^-$. This results in NV$^-$'s small optical cooperativity. Nevertheless, given the high finesse of the optical cavity, and the large magnetic dipole moment, the NV$^-$ can still be a candidate for the MO-NODE protocol if the small magnetic field is compatible with the experimental design.  The ${}^{117}\text{SnV}^-$ is favorable due to its zero-field protocol, and SiV$^0$ has potential to provide the NV$^-$-like magnetic characteristics and SnV$^-$-like optical characteristics provided that single-spin optical operation is possible.

\subsection{Perturbation Calculation of the Magnetic Dipole Moment of $^{117}\text{SnV}^-$}\label{Perturbation}

Following the analysis in \cite{harris2025high}, the unperturbed Hamiltonian contains strong spin-orbit and strain terms: 
\begin{equation}
    \Hamilo = \frac{1}{2}\lambda \Op{z}^L \Op{z}^S - \frac{1}{2} (\alpha_{E_{gx}} \Op{x}^L + \alpha_{E_{gy}} \Op{y}^L).
\end{equation}
The $\lambda$ is the spin-orbit coupling strength around $830$ GHz, and $\alpha$ is the strain along $x$ and $y$ direction. We add superscript $L$ and $S$ to denote the spin-1/2 operator of the electron and the orbit. The hyperfine interaction is treated as a perturbation in the Hamiltonian: 
\begin{equation}
    \Hamilp \approx \Hamil_{HF} = \frac{1}{4} A_{\parallel} \Op{z}^S \Op{z}^I + \frac{1}{4} A_{\perp} (\Op{x}^S \Op{x}^I + \Op{y}^S \Op{y}^I),
\end{equation}
where $\hat{\sigma}^I$ is the spin-1/2 operator of the nuclear spin. $A_{\parallel}$ and $A_{\perp}$ are the parallel and perpendicular hyperfine interaction strength. We first diagonalize the unperturbed Hamiltonian to find the unperturbed eigenstates. We define the splitting $\Delta = \sqrt{\lambda^2 + \alpha^2}$, where $\alpha = \sqrt{\alpha_{Egx}^2 + \alpha_{Egy}^2}$. The unperturbed system has two degenerate energy levels, $E^{(0)} = \pm \Delta/2$.

For clarity, we can define $ c_1 =\sqrt{\frac{\Delta + \lambda}{2\Delta}} , 
c_2 = \sqrt{\frac{\Delta - \lambda}{2\Delta}} $, and assume that $\alpha_{Egy}=0$ as implemented in \cite{harris2025high}. Then the four eigenstates can be written as 
\begin{align}
    \ket{\Psi_1} &= c_1 \ket{+L, \uparrow S} - c_2  \ket{-L, \uparrow S} \\
    \ket{\Psi_2} &= c_2 \ket{+L, \downarrow S} - c_1  \ket{-L, \downarrow S}
\end{align}
with energy $E^{(0)} = +\Delta/2$, and the other two 
\begin{align}
    \ket{\Psi_3} &= c_2  \ket{+L, \uparrow S} + c_1 \ket{-L, \uparrow S} \\
    \ket{\Psi_4} &= c_1  \ket{+L, \downarrow S} + c_2 \ket{-L, \downarrow S}
\end{align}
with energy $E^{(0)} = -\Delta/2$. We denote the two orbital branches as $\pm L$, and the two electron spin states as $\uparrow S$ and $\downarrow S$. To add in the hyperfine interaction, we use the degenerate perturbation by diagonalizing $\Hamilp$ within each degenerate manifold. We can look at the $E^{(0)} = -\Delta/2$ manifold, which has a $4 \times 4$ basis including nuclear spin:
$\{ \ket{\Psi_3, \up}, \ket{\Psi_3, \down}, \ket{\Psi_4, \up}, \ket{\Psi_4, \down} \}$. 

We notice that the states with aligned electron and nuclear spin $\ket{\Psi_3, \up}$ and $\ket{\Psi_4, \down}$ are already eigenstates with energy $\frac{A_{\parallel}}{4}$. So, we just need to diagonalize within the subspace $\{ \ket{\Psi_3, \down}, \ket{\Psi_4, \up} \}$. The Hamiltonian in this basis is 
\begin{equation}
    \hat{H}_{m_J=0} = 
    \begin{pmatrix}
        -A_{\parallel}/4 & \frac{A_{\perp} \alpha}{2\Delta} \\
        \frac{A_{\perp} \alpha}{2\Delta} & -A_{\parallel}/4
    \end{pmatrix},
\end{equation}
The eigenvalues are the energy corrections $E^{(1)}$: $E^{(1)} = -\frac{A_{\parallel}}{4} \pm \frac{A_{\perp} \alpha}{2\Delta}$. The corresponding eigenstates, which correspond to the $\ket{0}$ and $\ket{1}$ state in the main text are 
\begin{align}
    \ket{1}=\ket{E_{m_J=0'}^+} &= \frac{1}{\sqrt{2}} ( \ket{\Psi_3, \down} + \ket{\Psi_4, \up} ) \\
    \ket{0}=\ket{E_{m_J=0''}^+} &= \frac{1}{\sqrt{2}} ( \ket{\Psi_3, \down} - \ket{\Psi_4, \up} ).
\end{align}
Now, we can evaluate the matrix element $\bra{r}\vec{S}\ket{0} = \bra{\Psi_3, \up}\vec{S}\ket{0}$. The orbitals do not contribute to the magnetic dipole moment. The electron spins contribute to strain-dependent magnetic dipole of $\frac{1}{2\sqrt{2}}\frac{\gamma_e \alpha}{\Delta}$. The nuclear spin has a contribution three orders of magnitude smaller than the electron spin. Given the strain $\alpha = 928.4$ GHz in \cite{harris2025high}, the magnetic dipole of tin-vacancy is roughly $0.4$ of the NV$^-$ magnetic dipole. 

Our scheme may further be applied to molecular defects or other solid-state quantum emitters, which could have a stronger microwave coupling \cite{Rochman2023,wang2023single, Eichler2017, Haikka2017,Ranjan2020}; however, here we restrict the comparison to color centers. 

\begin{table*}
\begin{tabular}{|c|c|c|c|}
\hline  & NV$^-$\cite{Alkauskas2014} & Hyperfine ${}^{117}\text{SnV}^-$ \cite{harris2025high} & $\text{SiV}^0$\cite{Rose2018} \\
\hline Radiative Lifetime ($\tau_\text{rad}$) & 13 ns & 6 ns & 2 ns \\
\hline Debye-Waller Factor ($\xi$) & 0.03 & 0.6 & 0.9 \\
\hline ZPL Wavelength (nm) & 637 & 620 & 946 \\
\hline Selected ZPL Transition Dipole & 0.89 $D$ & 5.63 $D$ & 22.5 $D$ \\
\hline Optical Conditional Reflectivity & Yes & Yes & Unclear \\
\hline DC Magnetic Field $\vec{B}$ & 0.1-1 mT & 0 & 0.1-1 mT \\
\hline Engineered Strain $\alpha$ & 0 & 900 GHz & May Require \\
\hline MW qubit frequency (GHz)  & 2.9 & 0.6 & 1.0 \\
\hline Magnetic Dipole Moment & $\gamma_e/\sqrt{2}$ & $\alpha\gamma_e/(2\sqrt{2(\alpha^2+\lambda^2)})$ & $\gamma_e/\sqrt{2}$ \\
\hline Characteristic Impedance $Z$ & 0.12 $\Omega$ & 0.027 $\Omega$ & 0.042 $\Omega$\\
\hline Zero-Point Current Fluctuations $\delta I$ & 377 nA & 178 nA & 222 nA\\
\hline Capacitor Size (mm$^2$) & 0.95$\times$1.9 & 4.2$\times$8.5 & 2.7$\times$5.4\\
\hline
\end{tabular}
\caption{Comparison between different candidate color centers. $\gamma_e$ is the gyromagnetic ratio of the electron. The magnetic dipole of the tin-vacancy center depends on the strain $\alpha$ in the system, and the spin-orbit coupling $\lambda$. Here, $D=3.34\times10^{-30}\text{ C}\cdot\text{m}$. Color center parameters are used for calculating the optical cooperativity and spin microwave coupling strength. }\label{colortab}
\end{table*}

\subsection{Optical Cooperativity Estimation for NV$^{-}$}
\label{app:optical-cooperativity}

The small Debye-Waller factor of NV$^{-}$ makes emission
into the phonon sideband (PSB) an important competing channel
for its optical interface. To determine how this competition
affects spin-dependent reflection, we consider a three-level
model in which the cavity couples the ZPL transition
$\lvert g_1\rangle\leftrightarrow\lvert e\rangle$, while
$\lvert g_2\rangle$ represents an effective final manifold
for PSB emission and other unaddressed radiative transitions. Coupling between the
cavity and the PSB is neglected.

Let $\Gamma_1$ and $\Gamma_2$ be the background population
decay rates into the addressed ZPL and the remaining radiative channels, respectively. For a stable lower state, the optical
coherence decay rate  is
\begin{equation}
\gamma
=
\frac{\Gamma_1+\Gamma_2+\Gamma_{\rm nr}}{2}
+\gamma_\phi
=
\frac{1}{2T_1}+\gamma_\phi. 
\end{equation}
Each population-decay channel out of $\lvert e\rangle$
contributes half its rate to the decay of the optical
coherence. Consequently, PSB emission contributes to
$\gamma$ even though it does not couple to the cavity.

Writing
$\sigma_-=\lvert g_1\rangle\langle e\rvert$ and
$\sigma_z=\lvert e\rangle\langle e\rvert
-\lvert g_1\rangle\langle g_1\rvert$, the interaction
Hamiltonian is
$H_{\rm int}/\hbar=g(a^\dagger\sigma_-+a\sigma_+)$.
Within the rotating-wave and Markov approximations, the
Heisenberg-Langevin equations are
\begin{align}
&\dot a
=
-(i\omega_c+\kappa)a
-ig\sigma_-
+\sqrt{2\kappa_c}\,a_{\rm in}
+\sqrt{2\kappa_i}\,a_{i,{\rm in}},\\ \nonumber
&\dot\sigma_-
=
-(i\omega_a+\gamma)\sigma_-
+ig\sigma_z a.
\end{align}
Here $\kappa=\kappa_c+\kappa_i$ is the cavity
field-amplitude decay rate, $a_{\rm in}$ is the input
operator, and $a_{i,{\rm in}}$ describes
vacuum entering through internal-loss channels. The coupling $g$ is that of the physical ZPL
transition.

We calculate the linear response to a weak coherent probe
with the emitter initially prepared in $\lvert g_1\rangle$.
To first order in the probe amplitude, the excited-state
population and depletion of $\lvert g_1\rangle$ can be
neglected. Defining
$\alpha=\langle a\rangle$,
$s=\langle\sigma_-\rangle$, and
$\alpha_{\rm in}=\langle a_{\rm in}\rangle$, the expansion
about this prepared state gives
$\langle\sigma_z a\rangle=-\alpha$ to linear order.

Taking expectation values of the
equations therefore gives
\[
\begin{aligned}
\dot\alpha
&=-(i\omega_c+\kappa)\alpha-igs
  +\sqrt{2\kappa_c}\,\alpha_{\rm in},\\
\dot s
&=-(i\omega_a+\gamma)s-ig\alpha .
\end{aligned}
\]
Fourier transform the equations and introduce $\Delta_c=\omega_c-\omega$ with
$\Delta_a=\omega_a-\omega$. The solution is
\[
\begin{aligned}
s(\omega)
&=
-\frac{ig}{\gamma+i\Delta_a}\,\alpha(\omega),\\
\alpha(\omega)
&=
\frac{\sqrt{2\kappa_c}\,\alpha_{\rm in}(\omega)}
{\kappa+i\Delta_c+g^2/(\gamma+i\Delta_a)} .
\end{aligned}
\]
Using the input-output relation
$a_{\rm out}=a_{\rm in}-\sqrt{2\kappa_c}\,a$,
we obtain the reflection amplitude and the reflectivity:
\begin{equation}
\begin{aligned}
r(\omega)
&\equiv
\frac{\alpha_{\rm out}(\omega)}{\alpha_{\rm in}(\omega)}
=
1-\frac{2\kappa_c}{
\kappa+i\Delta_c+
\dfrac{g^2}{\gamma+i\Delta_a}
},\\
R(\omega)&=|r(\omega)|^2 .
\end{aligned}
\label{eq:optical-reflectivity}
\end{equation}

On resonance, $\omega=\omega_c=\omega_a$, the
emitter contributes $g^2/\gamma$ to the cavity response.
Its ratio to the bare cavity damping identifies the
optical cooperativity:
\begin{align}
C
&\equiv
\frac{g^2}{\kappa\gamma}
=
\frac{g_o^2}{\kappa_o\gamma_o},\\ \nonumber
r_{\rm res}
&=
1-\frac{2\kappa_c/\kappa}{1+C}.
\end{align}
With our definitions of the $g_o$ and $\gamma_o$, the Debye-Waller suppression is
already included in the physical ZPL coupling;
no additional division of the optical coherence decay
rate by the Debye-Waller factor is required.

At critical coupling, $\kappa_c=\kappa_i=\kappa/2$,
the empty cavity is nonreflective, whereas the coupled
emitter gives $r_{\rm bright}=C/(1+C)$.
Provided that the alternative logical spin state has
negligible optical susceptibility, it realizes the
empty-cavity response. For negligible pure dephasing and
a photon bandwidth over which the reflection amplitudes
are approximately constant, the reflectivities and
optical success probability reduce to
\begin{equation}
\begin{aligned}
R_{\rm bright}
&=\left(\frac{C}{1+C}\right)^2,
\qquad R_{\rm dark}=0,\\
P_{\rm opt}
&=\frac{1}{2}R_{\rm bright}
=\frac{1}{2}\left(\frac{C}{1+C}\right)^2 .
\end{aligned}
\label{eq:optical-success-probability}
\end{equation}
The factor $1/2$ accounts for the nonreflective logical
branch in the time-bin entanglement protocol.

\end{appendix}

\bibliography{reference}

\end{document}